\documentclass[aps,prb,twocolumn,showpacs]{revtex4}

\usepackage{graphicx} 
\usepackage{calc} 
\usepackage{amsfonts}  
\usepackage{amsmath} 

\begin{document}
\newcommand{\kw}{ \bar{k} }
\newcommand{\qq}{  \bar{q}_0  }
\newcommand{\qw}{ \bar{q} }
\newcommand{\G}{ \bar{G} }
\def\sumslashD{\mathop{\sum \kern-1.4em -\kern 0.5em}}
\def\sumslash{\mathop{\sum \kern-1.2em -\kern 0.5em}}
\def\intslash{\mathop{\int \kern-0.9em -\kern 0.5em}}
\def\intslashD{\mathop{\int \kern-1.1em -\kern 0.5em}}

\title{A renormalization group analysis of   the one-dimensional  extended Hubbard model }        
 
\author{M. M\'enard$^1$ and C. Bourbonnais$^{1,2}$}
\affiliation{$^1$ Regroupement Qu\'ebecois sur les Mat\'eriaux de Pointe,
  D\'epartement de physique, Universit\'e de Sherbrooke, Sherbrooke,
  Qu\'ebec, Canada, J1K-2R1 }
\affiliation{$^2$ Canadian Institute for Advanced Research, Toronto, Canada}
 
\date{\today}

\begin{abstract}
The phase diagram of the one-dimensional extended Hubbard model at half-filling is investigated by a weak coupling renormalization group method applicable beyond the usual continuum limit for the electron spectrum and coupling constants. We analyze  the influence of irrelevant momentum dependent interactions on asymptotic properties of the correlation functions and the nature of dominant phases for the lattice model under study.
  \end{abstract}
\pacs{}
\maketitle 

\section{Introduction}
 The application of the renormalization group (RG)  and boson representation methods to  one-dimensional (1D) models  of interacting electrons  have provided over the last four decades  considerable  insight into the nature of correlations in  low dimensional systems \cite{Emery79,Solyom79,Voit95,Giamarchi04}. This   has been largely achieved   by treating  the  models  in  the   continuum  field-theory  limit,   corresponding to the so-called weak coupling  1D electron gas (EG) model.  There are notable exceptions, however, where the 1D-EG model clearly fails to reveal the nature of correlations at long distance.  These situations are likely to occur in  models   for which lattice effects, albeit related  to irrelevant terms in the RG sense,  do affect the asymptotic behavior of  electronic correlations and then the nature of the ground state.
 
 

A well documented case is encountered in the  extended Hubbard    model at half-filling, which is  defined on a lattice in terms of intersite hopping and the on-site and nearest-neighbor sites couplings    $U$  and $V$. On numerical side, exact diagonalization\cite{Nakamura99,Nakamura00}, quantum Monte Carlo  \cite{Sandvik04} and density-matrix renormalization group  analysis  \cite{Zhang04} have established the incursion of a bond order-wave (BOW) state over a finite region of the phase diagram surrounding the   line  $U=2V>0$, a result   at variance   with the spin density-wave (SDW) to charge density-wave (CDW) transition  found  in  the theory of the 1D EG\cite{Emery79,Voit92,Japaridze99}. 

Using perturbation theory arguments,  Tsuchiizu and Furasaki\cite{Tsuchiizu02} showed how high-energy or short-distance degrees of freedom can modify the  initial conditions  of   an effective low energy continuum theory and  favor the occurrence of a  BOW phase that enfolds the  $U=2V>0$ line in weak coupling. The  influence of the lattice  on the nature of the ordered phase in this region of the phase diagram has been  investigated by Tam {\it  et al.,}\cite{Tam06} using the functional RG method.    The  scaling transformation of interactions, which in this framework  gather  both their marginal and  momentum dependent parts, was obtained  for  a  tight-binding  electron  spectrum in a finite momentum space. The  CDW/SDW degeneracy that takes place  at $U=2V$  in the continuum limit,  is thus lifted   and a BOW state stabilized over a portion of the  phase diagram that grows in size with increasing  $U$, consistently with numerical calculations at weak coupling.  The functional RG method, however,   tells us not as much about the structure of leading irrelevant terms and how these  modify  scaling  and  the nature of ordered states of the continuum theory.  

In this work we address this issue from a different perspective  that  generalizes  the weak coupling  momentum shell Kadanoff-Wilson (K-W) RG  method to lattice models \cite{Bourbon91,Bourbon03}. The proposed approach  exceeds the limitations of the continuum approximation and  takes into account the    tight-binding structure of the spectrum and its impact on the scaling transformation of  both local and momentum dependent interactions of the extended Hubbard model \cite{Dumoulin96b}. The latter couplings, though  irrelevant,   are found to affect the flow of the former interactions. A modification of     certain portions of the phase diagram follows;   in particular,   the  BOW phase  is found to insert  in  a finite region near the $U=2V>0$ line, in agreement with the results of numerical calculations.
 
 In Sec.~II, we introduce the model and  set out the basic steps of the momentum shell RG transformation for the partition function. In Sec.~III, the RG flow equations  for the coupling constants and the most singular response functions are analyzed at the one-loop level and different $U$ and $V$.  The phase diagram is mapped  out in weak coupling. 
We conclude in Sec. IV. 

\section{The extended Hubbard model and the renormalization group formulation}
\subsection{The model}
We consider the  extended Hubbard Hamiltonian  for a one-dimensional lattice, 
\begin{equation}
\label{ExHR}
\begin{split}
H= & -t\sum_{i,\sigma}(c^\dagger_{i+1,\sigma}c_{i,\sigma}+ c^\dagger_{i,\sigma}c_{i+1,\sigma})\cr
    &+ U\sum_{i}n_{i,\uparrow}n_{i,\downarrow} + V\sum_{i}n_in_{i+1}, 
    \end{split}
\end{equation}
where $t$ is is the hopping integral, $n_{i,\sigma}= c^{\dagger}_{i,\sigma} c_{i,\sigma}$  is the occupation number on site $i$ for the spin orientation $\sigma=\uparrow,\downarrow$, and $n_i= n_{i,\uparrow}+n_{i,\downarrow}$.   In Fourier space the Hamiltonian can be written  in the form 
\begin{widetext} 
\begin{equation}
\label{ExHF}
\begin{split}
 H= \sum_{p,k,\sigma} \epsilon_k c^\dagger_{p,k,\sigma}c_{p,k,\sigma} \ 
  & +  {1\over L}\sum_{\{k,q,\sigma\}} \left(g_1+ 2\bar{g}_1\sin^2 {q\over 2} \right) c^\dagger_{+,k_1+q +2k_F,\sigma_1}c^\dagger_{-,k_2-q-2k_F,\sigma_2}c_{+,k_2,\sigma_2}c_{-,k_1,\sigma_1}\cr
& +  {1\over L}\sum_{\{k,q,\sigma\}} \left(g_2+ 2\bar{g}_2\sin^2 {q\over 2} \right) c^\dagger_{+,k_1+q,\sigma_1}c^\dagger_{-,k_2-q,\sigma_2}c_{-,k_2,\sigma_2}c_{+,k_1,\sigma_1} \cr
& + {1\over 2L}\sum_{\{k,q,\sigma\}} \left(g_3+ 2\bar{g}_3\sin^2 {q\over 2} \right) \big(c^\dagger_{+,k_1+q+2k_F,\sigma_1}c^\dagger_{+,k_2-q -2k_F +G,\sigma_2}c_{-,k_2,\sigma_2}c_{-,k_1,\sigma_1} + \text{H.c.}\big)\cr
&+ {1\over 2L}\sum_{\{k,q,\sigma\}} \left(g_4+ 2\bar{g}_4\sin^2 {q \over 2} \right)  c^\dagger_{+,k_1+q,\sigma_1}c^\dagger_{+,k_2-q ,\sigma_2}c_{+,k_2,\sigma_2}c_{+,k_1,\sigma_1} \cr
& + {1\over 2L}\sum_{\{k,q,\sigma\}} \left(g_4+ 2\bar{g}_4\sin^2 {q \over 2} \right)  c^\dagger_{-,k_1+q,\sigma_1}c^\dagger_{-,k_2-q ,\sigma_2}c_{-,k_2,\sigma_2}c_{-,k_1,\sigma_1}, 
\end{split}
\end{equation}
\end{widetext}
where  $\epsilon_k = -2t\cos k$ is the tight-binding spectrum, and $v= 2t$ is the bare Fermi velocity;  the Fermi points are $k_F=\pm {\pi\over 2}$ at half-filling (here the lattice constant has been set to unity, and $\hbar= 1 = k_B)$. By analogy with the  `g-ology'  description   of    interactions, we have  proceeded  to the  splitting of  the $U$ and $V$ interaction terms  into couplings   for   right ($p=+$, $k>0$) and left ($p=-$, $k<0$) moving electrons. We thus obtain momentum independent (local) as well as  momentum dependent (non local)  couplings,    denoted  in (\ref{ExHF}) by  $g_{i=1\ldots4}$ and $\bar{g}_{1\ldots4}$, respectively. The  pairs of couplings for backscattering ($g_1,\bar{g}_1$) and Umklapp ($g_3,\bar{g}_3$) have the bare amplitudes  $g_{1,3}= U-2V, \bar{g}_{1,3}= 2V$, whereas $ g_{2,4}=  U +2V$ and $\bar{g}_{2,4}=-2V$ stand  for the amplitudes for the  forward scattering between opposite  ($g_2,\bar{g}_2$) and parallel  ($g_4,\bar{g}_4$) $k$ electrons. 

The information about the lattice in (\ref{ExHF}) is present by the use of   the tight binding spectrum $\epsilon_k$ for $k\in [-\pi,\pi]$ in the Brillouin zone and in the momentum dependent  couplings $\bar{g}_i\sin^2 q/2$. In the continuum limit, the latter amplitudes vanish when evaluated at zero momentum transfer, while the  spectrum  $\epsilon_k\to \epsilon_p(k) \approx v(pk-k_F)$  is taken as  linear  around each Fermi points. One thus  recovers    the standard electron gas formulation of the extended Hubbard model \cite{Emery79,Solyom79,Voit95,Giamarchi04}. 

\subsection{The renormalization group transformation}
We write the partition function $Z=\text{Tr}\, e^{-\beta H} $    as a functional integral \begin{equation}
\label{Z}
Z= \int\!\!\int\mathfrak{D}\psi^*\mathfrak{D}\psi\, e^{S[\psi^*,\psi]},
\end{equation}
over anticommuting Grassmann  fields $\psi^{(*)}$.   The action $S[\psi^*,\psi] =S_0[\psi^*,\psi] + S_{I}[\psi^*,\psi]$ consists of a  free and an  interacting  parts. In the Fourier-Matsubara space, the former part  $S_0[\psi^*,\psi]$ reads
\begin{equation}
\label{S0}
S_0[\psi^*,\psi]= \sum_{p,\bar{k},\sigma} [G^{0}_p(\bar{k})]^{-1}\psi^*_{p,\sigma}(\bar{k})\psi_{p,\sigma}(\bar{k}),
\end{equation}
where 
\begin{equation}
\label{G0}
G^{0}_p(\bar{k}) = [i\omega_n - \epsilon_k]^{-1},
\end{equation}
is the free electron propagator. Here  $\kw = (k,\omega_n)$ and $\omega_n=(2n+1)\pi T$ is the fermion Mastubara frequency. The interacting part is given by 
\begin{widetext}
\begin{equation}
\begin{split}
S_I[\psi^*,\psi] =  & -{T\over L}  \sum_{\{\bar{k},\bar{q},\sigma\}} \left(g_1+ 2\bar{g}_1\sin^2 {q\over 2} \right) \psi^*_{+,\sigma_1}(\kw_1 +\qq +\qw )\psi^*_{-,\sigma_2}(\kw_2 -\qq -\qw )\psi_{+,\sigma_2}(\kw_2 )\psi_{-,\sigma_1}(\kw_1) \cr
  &-{T\over L}  \sum_{\{\bar{k},\bar{q},\sigma\}} \left(g_2+ 2\bar{g}_2\sin^2 {q\over 2} \right) \psi^*_{+,\sigma_1}(\kw_1   +\qw )\psi^*_{-,\sigma_2}(\kw_2  -\qw )\psi_{-,\sigma_2}(\kw_2 )\psi_{+,\sigma_1}(\kw_1)\cr
 & - {T\over 2 L}  \sum_{\{\bar{k},\bar{q},\sigma\}} \left(g_3+ 2\bar{g}_3\sin^2 {q\over 2} \right) \big(\psi^*_{+,\sigma_1}(\kw_1   +\qq +\qw )\psi^*_{+,\sigma_2}(\kw_2 -\qq -\qw + \G )\psi_{-,\sigma_2}(\kw_2 )\psi_{-,\sigma_1}(\kw_1) + \text{c.c.}\big)\cr
 & - {T\over 2 L} \sum_{\{\bar{k},\bar{q},\sigma\}} \left(g_4+ 2\bar{g}_4\sin^2 {q\over 2} \right) \psi^*_{+,\sigma_1}(\kw_1   +\qw )\psi^*_{+,\sigma_2}(\kw_2  -\qw )\psi_{+,\sigma_2}(\kw_2 )\psi_{+,\sigma_1}(\kw_1) \cr
 &- {T\over 2 L} \sum_{\{\bar{k},\bar{q},\sigma\}} \left(g_4+ 2\bar{g}_4\sin^2 {q\over 2} \right) \psi^*_{-,\sigma_1}(\kw_1   +\qw )\psi^*_{-,\sigma_2}(\kw_2  -\qw )\psi_{-,\sigma_2}(\kw_2 )\psi_{-,\sigma_1}(\kw_1),
\end{split}
\end{equation}
\end{widetext}
where $\qw=(q,\omega_m)$, $\omega_m=2\pi mT$, $\qq=(2k_F,0)$; here $\G=(4k_F,0)$ is a reciprocal lattice vector that enters in the definition of Umklapp scattering at half-filling.
 
 The momentum shell K-W RG transformation is based upon the recursive application of the  two following steps for the partition function. In the first step, a partial trace of $Z$ over   outer shell electronic degrees of freedom  denoted by $\bar{\psi}_{p,\sigma}(k,\omega_n)$, is carried out  at all $\omega_n$ and spin $\sigma$. The outer momentum shell is defined by the intervals of momentum 
 \begin{align}
\label{shell}
k \in \  & [0,k_F-k_0/s[ \ \cup \  ]k_F+k_0/s,\pi], \ \  p=+ \cr
     \in \   & ]\!-k_F+k_0/s,0] \ \cup \   [-\pi,-k_F-k_0/s[, \ \  p=-. \cr
\end{align} 
 above and below the Fermi level   for each branch $p$.  Here $k_0=\pi/2$ is a cutoff wave vector   defined with respect  to the Fermi points (\hbox{$\pm k_F\pm k_0=\pm \pi$}), and $s=e^{d\ell}>1$ is the momentum scaling factor  for  $d\ell\ll 1$. The second step consists in the rescaling of the momentum distance from the Fermi points we call  $\delta k$; this gives  $k'= \pm k_F + s \delta k $, which restores the initial cutoff $k_0=\pi/2$ of the lattice model.

 The two recursive steps of the RG transformation can be expressed as
 \begin{align}
\label{RG}
  & Z   =\! \Big[\int\!\!\int_<\mathfrak{D}\psi^*\mathfrak{D}\,\psi e^{S[\psi^*,\psi]_\ell} \!\!  \int\!\!\int  \mathfrak{D}\bar{\psi}^*\mathfrak{D}  \bar{\psi}  \,e^{ S_0[\bar{\psi}^*,\bar{\psi}] }\cr
    &\hskip 4 truecm \ \ \times e^{ \sum_{i=1}^4 S_{I,i}[\bar{\psi}^*,\bar{\psi},\psi^*,\psi]}\Big]_{\psi\to\zeta^{1\over 2}_s \psi'} \cr
   &\propto\!\! \Big[\!\ \! \int\!\!\int_<\! \mathfrak{D}\psi^*\mathfrak{D}\psi \ e^{S[\psi^*,\psi]_\ell + \langle S_{I,2}\rangle_{\bar{0},c} +\frac{1}{2} \langle (S_{I,2})^2\rangle_{\bar{0},c} +\ldots }\Big]_{\psi\to\zeta^{1\over 2}_s \psi'},\cr
\end{align} 
where $S_{I,i}$ is the interacting part of the action with $i=1,\ldots,4$,  $\bar{\psi}$ fields in the outer momentum shell. The outer shell statistical averages $\langle ....\rangle_{\bar{0},c}$ over the variables $\bar{\psi}^{(*)} $ are performed with respect to $S_0[\bar{\psi}^*,\bar{\psi}]$. These averages correspond to the sum of all connected diagrams with even number of external fields $\psi^*,\psi$ pertaining to  the inner  momentum shell ($<$)   degrees of freedom, which  are kept fixed in the partial trace operation. At the one-loop level  the partial trace and rescaling lead  to the recursion relations 
 \begin{align}
\label{RecurS0}
    \Big[S_0[\psi^*,\psi]_{\ell + d\ell}  =  &\ S_0[\psi^*,\psi]_\ell + \langle S_{I,2}\rangle_{\bar{0},c} + \ldots \Big]_{\psi\to\zeta^{1\over 2}_s \psi'}, \\
    \label{RecurSI}
      \Big[S_I[\psi^*,\psi]_{\ell + d\ell}  =  &\ S_I[\psi^*,\psi]_\ell + \frac{1}{2} \langle S^2_{I,2}\rangle_{\bar{0},c} + \ldots \Big]_{\psi\to\zeta^{1\over 2}_s \psi'}, 
\end{align}
for the free and interacting parts of the action. 

Following the momentum  rescaling, the inner shell fields $\psi$ are  rescaled by the factor $\zeta_s^{1/2}$, which can be derived from a dimensional analysis of the parameters that define the bare action $S_0$.  Thus assuming that  the rescaling of  the tight-binding spectrum is of   the form $\epsilon'_{k'}\equiv \zeta_s \epsilon_k $, by taking $k'= \pm k_F + s\delta k$, one gets  
 in the limit $d\ell \to0$ 
\begin{equation}
\zeta_s \to \zeta_s(\delta k)= s^{\delta k\cot \delta k},
\end{equation}
which can   be expressed in the form $s^y$ compatible with an iterative transformation in renormalization group. At variance with the usual case, however,  the scaling dimension $y$ is here $k$ dependent. Thus at either the  edge or the  bottom of the band where the group velocity  vanishes, $\zeta_s(\pm k_0) \to s^0$ and $\epsilon_k$ is dimensionless. It  is only when the Fermi points  is  approached in the limit $\delta k\to 0$,  that $\zeta_s(\delta k) \to s^1$ and  the result of the continuum limit for a linear spectrum is recovered\cite{Bourbon91}.  This also indicates that repetition of rescaling   turns down the curvature of the band, which continuously evolves toward a linear shape.   Since $\omega_n$ or the temperature $T$ enters on the same footing as   $\epsilon_k$ in the inverse propagator $[G^0]^{-1}$, the   temperature   then transforms  according to   $T'=\zeta_s(\delta k)T$. Now referring to the form of $S_0$ in (\ref{S0}), this   yields the  transformation  assumed   above for the field, namely $\psi^{(*)\prime}=\zeta^{-1/2}_s(\delta k) \psi^{(*)}$.
 
When applied to the interacting part $S_I$ of the the action, the above  relations for the field and temperature,  combined to   the shrinking of the number sites $L'=L/s$ under rescaling,    will impose   the following $k$-dependent transformations of interactions
 \begin{align}
\label{recura}
   g'_i & = \big(g_i +{\cal O}(g^2) \big) s^{-1+ \delta k \cot  \delta k}, \\
    \bar{g}'_i& =  \big( \bar{g}_i + {\cal O}(g\bar{g}) \big) s^{-1-\delta k \csc \delta k}.
    \label{recurb}
\end{align} 
  It ensues that for $\delta k\to \pm k_0$, we have  $g_i'\to g_i s^{-1}$, and the local couplings are then irrelevant instead of being marginal  variables near  the bottom or the edge of the band. At the approach of the Fermi level, when $\delta k\to 0$, we have    $g_i'=g_i$ and  the  dimensionless or marginal character of the local interactions of the electron gas model is retrieved\cite{Solyom79,Bourbon91}. In the same way, the non local terms transform  according to $ \bar{g}'_i  =   \bar{g}_i s^{-1-\pi/2}$ at the boundaries or the bottom of the band and are therefore strongly irrelevant. In the limit $\delta k \to 0$ near the Fermi points,  $ \bar{g}'_i  =   \bar{g}_i s^{-2}$, which corresponds to the usual negative bare scaling dimension of nearest-neighbor  couplings  of  the continuum theory\cite{Haldane82,Giamarchi04}. 
  
  \subsection{The Fermi velocity and coupling  constant flow equations}
  We  now proceed to the partial trace  operation  that defines   the  first step of the renormalization group transformation  (\ref{RG}). At the one-loop level, this amounts to evaluate the outer shell statistical averages  $\langle S_{I,2}\rangle_{\bar{0},c}$ and $\langle S^2_{I,2}\rangle_{\bar{0},c}$ of the recursion relations (\ref{RecurS0}) and (\ref{RecurSI}). The former contribution $\langle S_{I,2}\rangle_{\bar{0},c}$ is composed of Hartree and Fock self-energy corrections. In these, enter   $k$ independent or constant  terms that correct the chemical potential, a quantity that can be simply redefined to  keep the filling of the band constant.  These   terms can be safely ignored. The presence of non local interactions give rise to momentum dependent Fock terms, which at the step $\ell$ of the iterative RG procedure read  
\begin{align}
\label{HF}
\langle S_{I,2}\rangle_{\bar{0},c} =   {T(\ell)\over L(\ell)}\sum_{\bar{k}}\sumslashD_{\bar{k}'}& [\, \bar{g}_1(\ell)G_{-p}^0(\bar{k}') 
- \bar{g}_4(\ell)G_{p}^0(\bar{k}')] \cr 
&\times \cos(k_F + \delta k)  \psi^*_p(\bar{k})\psi_p(\bar{k}),
\end{align}
where the slashed summation contains an integration over $k'$ in the outer momentum shell interval (\ref{shell}) at a given $p$.  The Fock terms contribute to the renormalization of the spectrum, that is   the  Fermi velocity. Carrying the $\bar{k}'$ summation, one gets the flow equation for the velocity,  
  \begin{equation}
  \label{velocity}
 {d_\ell \ln v(\ell)} =  {\pi \over 4} \big(\tilde{\bar{g}}_4(\ell) -\tilde{\bar{g}}_1(\ell)\big) \tanh[ v(\ell) \sin \delta k_\ell/2T],  
\end{equation}
where $\delta k_\ell  = k_0e^{-\ell}$ and the couplings  $\tilde{\bar{g}}\equiv \bar{g}/\pi v(\ell)$ are henceforth taken as  normalized by the scale dependent Fermi velocity $v(\ell)$.

The recursion relations (\ref{recura}) for the local normalized couplings  $\tilde{g} (\equiv{g}/\pi v(\ell))$ are obtained from the  outer  shell contractions   $\langle S^2_{I,2}\rangle_{\bar{0},c}$  in  the logarithmically  singular Cooper (electron-electron) and Peierls ($2k_F$ electron-hole) channels. Their insertion in (\ref{recura}), leads after rescaling to the recursion relations  
\begin{align}
\label{}
  \tilde{g}_1' = &\  [\, \tilde{g}_1+ ( -\tilde{g}_1^2 +\tilde{g}_3\tilde{\bar{g}}_3 ) I_P + \tilde{\bar{g}}_1(\tilde{g}_2 +\tilde{\bar{g}}_2) I_C\, ]s^{-f_g}      \\
  \tilde{g}_2' = &\  [ \, \tilde{g}_2+ ( \tilde{g}_1+  \tilde{\bar{g}}_1)^2 I_C  + ( \tilde{g}_3 +  \tilde{\bar{g}}_3)^2 I_P\,]   s^{-f_g} \\  
    \tilde{g}_3' = &  \ [ \, \tilde{g}_3+ ( \tilde{g}_2+  \tilde{\bar{g}}_2) (2 \tilde{g}_3+  \tilde{\bar{g}}_3)I_P  \cr
   & \hskip 1 truecm - ( \tilde{g}_1+  \tilde{\bar{g}}_1)( \tilde{g}_3- \tilde{\bar{g}}_3)I_P\, ] s^{-f_g} \\
  \tilde{g}_4' = &  \  \tilde{g}_4 s^{-f_g}, 
\end{align}
where   $f_g=  1-\delta k_\ell \cot \delta k_\ell  + d_\ell \ln v(\ell) $, which contains the rescaling exponent  of (\ref{recura}),  and the correction due to the normalization from the   scale dependent Fermi  velocity. We note that the one-loop level, there is no logarithmic correction to the forward scattering amplitude $g_4$.  The outer shell  Cooper and Peierls loops  evaluated at zero external  variables  are respectively given by  
 \begin{align}
\label{ }
I_C=  &-  2\pi v(\ell) {T(\ell) \over L(\ell)}\sumslashD_{k>0} \sum_{\omega_n} G^0_+(\bar{k}+\bar{q}_C)G^0_-(-\bar{k})\cr
     = &- \pi v(\ell) {1\over L(\ell)} \sumslashD_{k>0}  {\tanh [\epsilon(k)/2T(\ell)]\over \epsilon({k})}\cr
= & - {\pi \over 2} \tanh [\epsilon(\ell)/2T] \ d\ell
\end{align}
 at $\bar{q}_C=0$, where $\epsilon(\ell) = v(\ell)\sin\delta k_\ell$   and 
  \begin{align}
\label{ }
I_P(\ell) =  &-  2\pi v(\ell) {T(\ell) \over L(\ell)}\sumslashD_k \sum_{\omega_n} G^0_+(\bar{k}+\bar{q}_P)G^0_-(k)\cr
= &  -I_C
\end{align}
at $\bar{q}_P= (2k_F,0)$. It is worth stressing that neglecting the dependence of $I_{P,C}$ on external variables does not generate new momentum dependent interactions whose number is kept fixed along the RG flow.

 The flow equations for the local interactions then become
  \begin{align}
\label{flowg1}
 d_\ell \tilde{g}_1   = & - f_g \tilde{g}_1 + f_1[-\tilde{g}_1^2 -\tilde{\bar{g}}_1(\tilde{g}_2+\tilde{\bar{g}}_2) + \tilde{g}_3\tilde{\bar{g}}_3\, ],\\
 \label{flowg2}
d_\ell \tilde{g}_2    = &  - f_g  \tilde{g}_2 + {1\over 2} f_1[ \,(\tilde{g}_3+\tilde{\bar{g}}_3)^2 - (\tilde{g}_1+\tilde{\bar{g}}_1)^2], \\
\label{flowg3}
d_\ell \tilde{g}_3   = &   - f_g  \tilde{g}_3  +  f_1  [ \, (\tilde{g}_2+ \tilde{\bar{g}}_2)(2\tilde{g}_3+\tilde{\bar{g}}_3)\cr
                      & \hskip 2.6 truecm - (\tilde{g}_1+\tilde{\bar{g}}_1)(\tilde{g}_3-\tilde{\bar{g}}_3)],  \\
                      \label{flowg4}
d_\ell \tilde{g}_4  = &   - f_g  \tilde{g}_4, 
\end{align}
where   $f_1= {\pi\over 2}  \tanh [\epsilon(\ell)/2T ]$. These equations differs from the usual scaling equations of the 1D-EG model in two respects. First, the   rescaling for a tight-binding spectrum  and velocity renormalization introduce   linear terms; second, there are additional corrections coming to the coupling to  momentum dependent interactions. These latter corrections are by far the most likely to influence the flow of local couplings if not the nature of the ground state as we will see.

As for the non local irrelevant interactions, the  corrections due to loop  contractions are small and  will be  neglected in weak coupling.  From the rescaling transformation (\ref{recurb}) and  the normalization of the couplings by $\pi v(\ell)$, we get 
 \begin{align}
\label{gbar}
  d_\ell \tilde{\bar{g}}_{i}    = &- \big(1+\delta k_\ell \csc \delta k_\ell + d_\ell \ln v(\ell)\big) \tilde{\bar{g}}_{i} 
  \end{align}  
for $i=1,\ldots 4$.  In the zero temperature limit, the solution of Eqs. (\ref{velocity}) and (\ref{gbar}) yields the following expressions 
\begin{equation}
\label{velocityb}
v(\ell) = v\left(1- {V\over \pi t}\ln[ 2\cos^2(\delta k_\ell/2)]\right),
\end{equation}
for the Fermi velocity and  
\begin{equation}
\label{gbarb}
{\bar{g}}_{i}(\ell) = {\bar{g}}_{i} {v\over v(\ell)} e^{-\ell}\tan (\delta k_\ell/2),
\end{equation}
for the non-local couplings. The Fermi velocity is thus renormalized downward  due to the presence of the $V$ term; it  reaches the value $v^*=v(1-{V\over\pi t}\ln2)$  in the limit of large $\ell$.
\subsection{Response Functions}
To determine the nature of long-range correlations in the ground state, we consider the most  singular response functions or susceptibilities, which are denoted  $\chi_\mu$.   
These latter are obtained by adding to  the $\ell=0$ action an additional term $S_h$\cite{Bourbon91}, which consists of   source fields $h_\mu$   linearly coupled to the composite fields $O_\mu^*$,
\begin{equation}
\label{ }
S_h[\psi^*,\psi] =  \sum_{\mu,\bar{q}} [\, h_\mu(\bar{q}) z_\mu O_\mu^*(\bar{q}) + {\rm c.c}\,],
\end{equation}
where $z_\mu$ is a pair vertex renormalization factor ($z_\mu=1$ at $\ell=0$).  In what follows we shall examine the site spin density-wave ($\mu=$ SDW), bond spin density-wave ($\mu=$ BSDW), site charge density-wave ($\mu=$ CDW) and  the bond order-wave ($\mu=$ BOW) susceptibilities of the Peierls channel;   the singlet (SS) and triplet (TS) superconducting susceptibilities of the Cooper channel. These are defined with the aid of the following expressions for the composite pair fields,
\begin{align}
\label{ }
O_{{\rm SDW/BSDW}}(\bar{q}) =  &  {1\over2} \big(O^*_{x,y,z}(-\bar{q}) \pm O_{x,y,z}(\bar{q}) \big)
\end{align}
 and 
 \begin{align}
\label{ }
O_{{\rm CDW/BOW}}(\bar{q}) =  & {1\over2} \big(O^*_{0}(-\bar{q}) \pm O_{0}(\bar{q})\big)  
\end{align}
in the Peierls channel, where  
$$
O_{\mu}(\bar{q}) =  \sqrt{T\over L}  \sum_{\bar{k}} \psi(\bar{k}-\bar{q})^*_{-,\alpha}\sigma_\mu^{\alpha\beta}  \psi_{+,\beta}(\bar{k}),
$$  
 and 
\begin{align}
\label{ }
O_{{\rm SS}}(\bar{q}) =  & \sqrt{T\over L}  \sum_{\bar{k}} \alpha \psi(-\bar{k}+\bar{q})^*_{-,-\alpha}  \psi_{+,\alpha}(\bar{k}),  \\
O_{{\rm TS}_{\mu}}(\bar{q}) = &  \sqrt{T\over L}  \sum_{\bar{k}} \alpha \psi(-\bar{k}+\bar{q})^*_{-,-\alpha}\sigma_\mu^{\alpha\beta}  \psi_{+,\beta}(\bar{k})
\end{align}
in the Cooper channel. Here $\sigma_{\mu=x,y,z} (\sigma_0)$ are the Pauli (identity) matrices.

The renormalization group transformation (\ref{RG})  at the one-loop level, will modify $S_h$ according to   
\begin{align}
\label{h}
S_h[O^*,O]_{\ell +d\ell} =\!  \Big[ S_h[O^*,O]_{\ell}  &+  \langle S_h S_{I,2} \rangle_{\bar{0},c} + \ldots \Big]_{O^{(*)} \to s^0  O'^{(*)}}\cr
                                                                                & + {1\over 2}  \langle S^2_h \rangle_{\bar{0},c} + \ldots,
\end{align}
where the pair fields, having zero canonical dimension, remain unchanged under rescaling. The last term is a constant  $\propto d\ell z_\mu^2 h_\mu^*h_\mu$ that  adds at each iteration and yields the expression of the susceptibility 
\begin{equation}
\label{ }
 \pi v\chi_\mu(\bar{q}^0_\mu)= {\pi\over 2} \int_\ell {v\over v(\ell)}z_\mu^2  \tanh [\epsilon(\ell)/2T] d\ell,
\end{equation}
which is defined positive and evaluated  in the static limit at $\bar{q}^0_\mu =(2k_F,0)$ and $(0,0)$ for  the Peierls and Cooper channels, respectively.  From the one-loop outer shell corrections to the linear coupling, which read  
$$
\langle S_h S_{I,2} \rangle_{\bar{0},c} ={\pi\over 2}  \tanh [\epsilon(\ell)/2T]d\ell  \sum_{\mu, \bar{q}} [\, h_\mu(\bar{q}) \tilde{g}_\mu z_\mu O_\mu^*(\bar{q}) + {\rm c.c}\,],
$$
one gets the one-loop   equation  for the pair vertex part $z_\mu$ at $\bar{q}_\mu^0$,
\begin{equation}
\label{zmu}
{d_\ell \ln z_\mu} = \tilde{g}_\mu {\pi\over 2} \tanh [\epsilon(\ell)/2T]. 
\end{equation}
For the density-wave type susceptibilities, the  normalized couplings $\tilde{g}_\mu$ are given by the combinations  
\begin{align}
\label{gmuP}
 \tilde{g}_{\rm CDW/BOW}   &  =  -2\tilde{g}_1+ \tilde{g}_2  + \tilde{\bar{g}}_2 \mp \tilde{g}_3 \pm  \tilde{\bar{g}}_3,   \\
  \tilde{g}_{\rm SDW/BSDW} & = \tilde{g}_2\pm \tilde{g}_3 + \tilde{\bar{g}}_2 \pm  \tilde{\bar{g}}_3. 
 \end{align}
 The corresponding expressions for the superconducting susceptibilities are 
 \begin{align}
\label{gmuC}
 \tilde{g}_{\rm SS/TS}   &  =  \mp \tilde{g}_1-\tilde{g}_2 \mp \tilde{\bar{g}}_1 - \tilde{\bar{g}}_2.   
 \end{align}  
A positive value for $\tilde{g}_\mu$ at $\ell \to \infty$ signals a singularity in $z_\mu$ and then in $\chi_\mu$ in that limit.
 
  \section{Results}
  
  \begin{figure}
 \includegraphics[width=7.0cm]{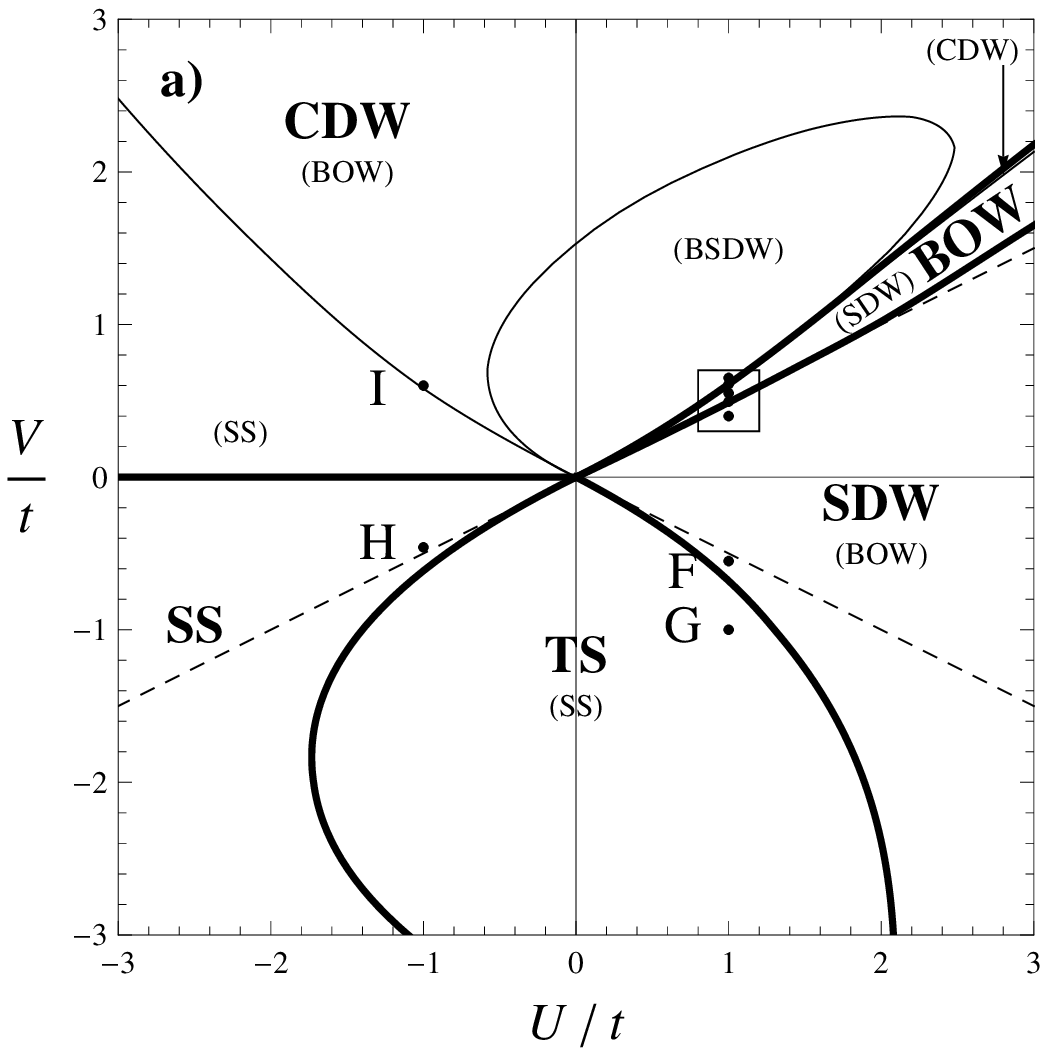} \includegraphics[width=7.0cm]{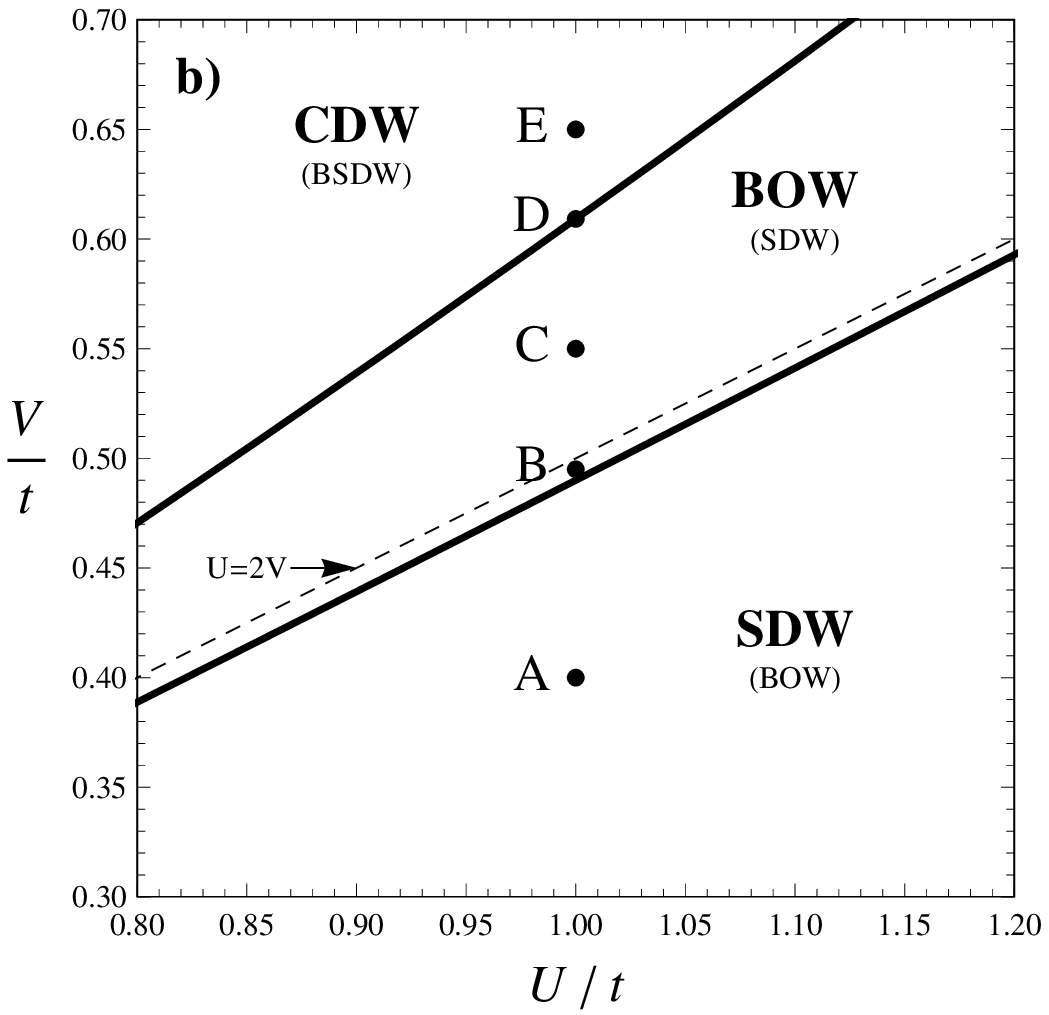}
 \caption{ a) The phase diagram of the 1D extended Hubbard model.  The bold (thin) lines refer to the boundaries between  the primary (secondary) phases indicated in bold (regular) characters. The dashed lines correspond 	to the boundaries of the phase diagram of the electron gas model in the continuum limit; b)  zoom in the neighborhood of the $U=2V$ (dashed) line in the repulsive sector. \label{Dphases} 
  } 
 \end{figure}

The  solution of the flow equations for the pair vertices (\ref{zmu}) and the couplings (\ref{flowg1}-\ref{gbar}) in the $T\to 0$ limit  leads to the determination of the most   singular  susceptibilities. These  in turn serve  to the determination   of the  dominant and subdominant phases of the model in the ground state. This is summarized in the one-loop phase diagram of Fig.~\ref{Dphases},  as a function of weak $U$ and $V$. The results are  compared with those obtained in the continuum  limit \cite{Emery79,Giamarchi04}.   
 
 \subsection{Repulsive  U }
\begin{figure}
 \includegraphics[width=7.0cm]{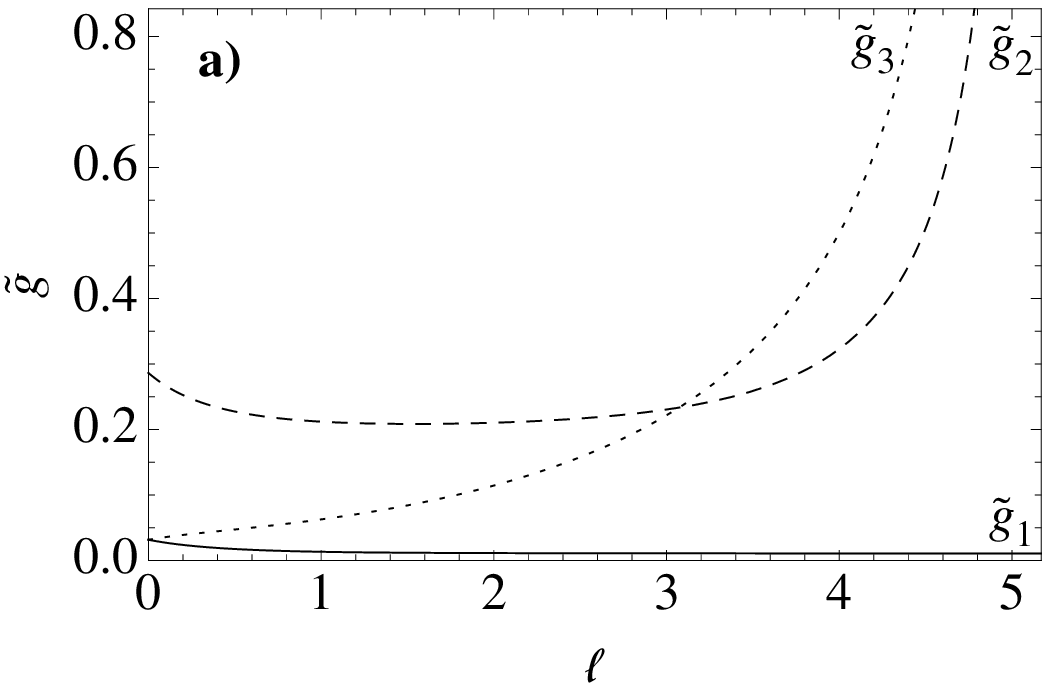}
 \par
 \includegraphics[width=7.0cm]{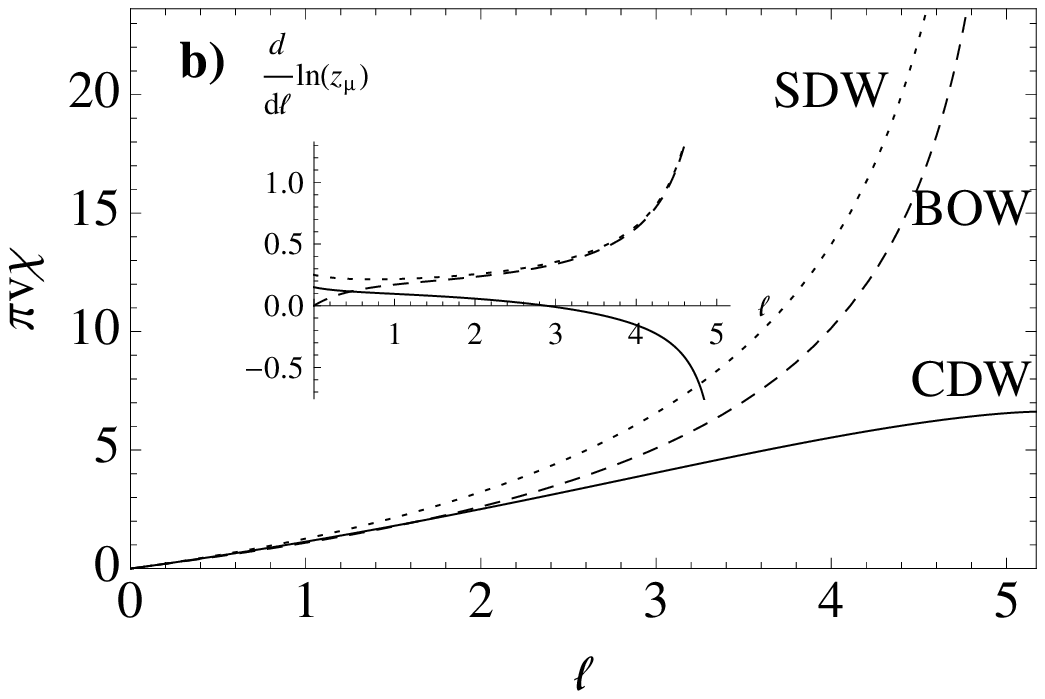}
 \caption{ a)  Flow of the coupling constants $\tilde{g}_{1,2,3}$ at the point A (1, 0.4)  of the phase diagram in Fig.~\ref{Dphases}. b) The density-wave susceptibilities {\it vs} $\ell$; inset: the flow of the pair vertices $d_\ell \ln z_\mu$  for $\mu=$ SDW, BOW and SDW.  \label{A}} 
 \end{figure}

 We commence by looking at the first quadrant of the phase diagram, in the region surrounding the $U=2V>0$ line.  At the point A  below the separatrix  in Fig.~\ref{Dphases}-b, where $U>2V$, the  $\tilde{g}_2$ and $\tilde{g}_3$ couplings  scale to strong repulsive values and become singular at a finite $\ell_\rho$,  a singularity at one-loop level that is   indicative of a (Mott) gap in the charge sector compatible with the initial conditions satisfying the inequality $\tilde{g}_1-2\tilde{g}_2 < \tilde{g}_3$. The repulsive $\tilde{g}_1$ coupling is marginally irrelevant and attributed to   gapless spin degrees of freedom. The SDW response then develops a  singularity similar to the one  of  BOW   at large  $\ell\sim\ell_\rho$,  as shown by the behavior of $z_{\rm SDW}$ and $z_{\rm BOW}$  in the inset of Fig.~\ref{A}-b. 
From the  same Figure, however, the  amplitude of the SDW susceptibility is larger, and  SDW (BOW) is then taken as  the dominant (subdominant)  phase in the ground state. These one-loop results indicate that in this region irrelevant non local couplings introduce no qualitative changes  with respect to known results of the continuum theory\cite{Kimura75,Emery79,Solyom79}.  
   \begin{figure}
 \includegraphics[width=7.0cm]{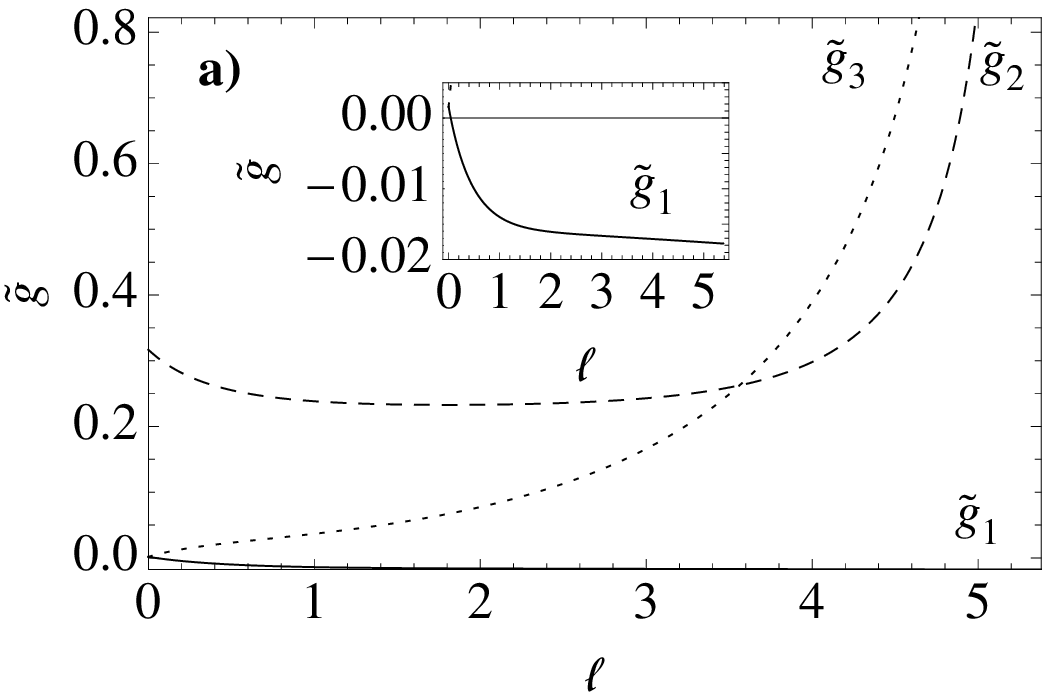} \includegraphics[width=7.0cm]{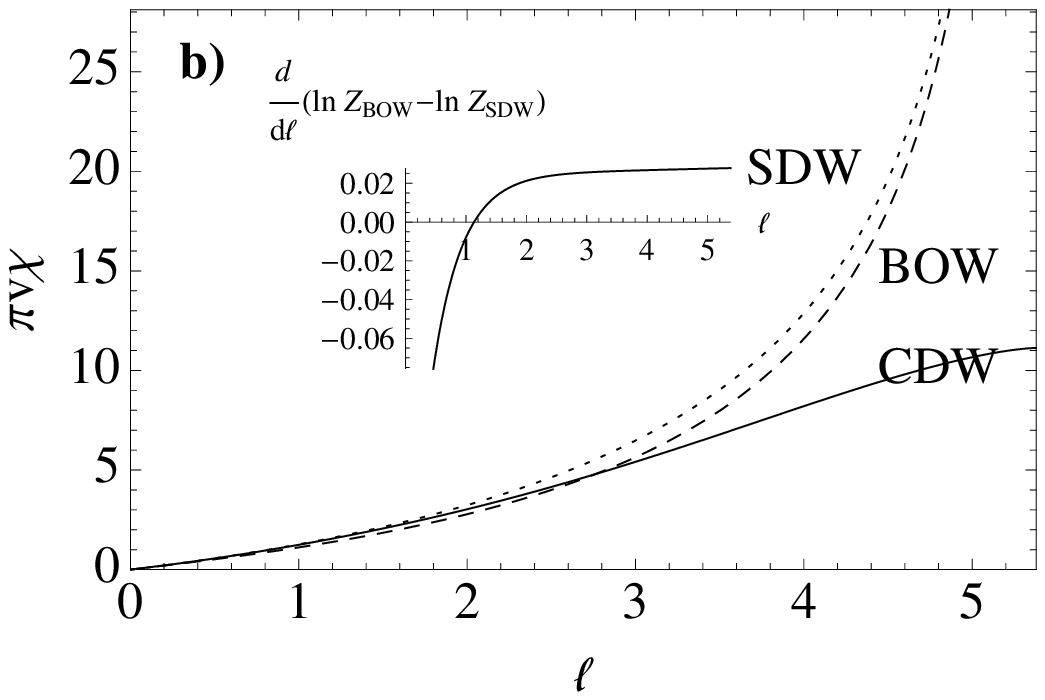}
 \caption{ a)  Flow of the coupling constants $\tilde{g}_{1,2,3}$ at point  B : (1, 0.495)  of the phase diagram in Fig.~\ref{Dphases}. b) The density-wave susceptibilities {\it vs} $\ell$; inset: the difference  between  the BOW and SDW flows of the pair vertices showing the dominance of the BOW phase.  \label{B}} 
 \end{figure}

If we now move up  to the point B in the phase diagram of Fig.~\ref{Dphases}-b,  close  but below the $U=2V$ line,  a qualitative change with respect to the results of the continuum limit emerges. While $\tilde{g}_2$ and $\tilde{g}_3$ still scale  to strong repulsive coupling,  signaling the formation of a charge gap at a finite $\ell_\rho$ (Fig.~\ref{B}-a), the backscattering amplitude $\tilde{g}_1$ no longer scales toward zero, but extends across the $\tilde{g}_1=0$  line to then level  off at a small   non universal negative value (inset of Fig.~\ref{B}-a). According to the expressions in  (\ref{gmuP}),  this change of sign of $\tilde{g}_1$  yields   $\tilde{g}_{\rm BOW} > \tilde{g}_{\rm SDW}$,   indicating that  the strongest singularity  now occurs for the  BOW  response (inset of Fig.~\ref{B}-b). The BOW  phase then becomes the dominant  phase, whereas SDW closely follows as the secondary phase. The   change of sign of $\tilde{g}_1$ takes its origin in the presence of non local couplings in the   flow equations (\ref{flowg1}-\ref{flowg3}). Although irrelevant, these interactions   push the renormalization of $\tilde{g}_1$ ($\tilde{g}_3$) downward (upward) through their coupling to local variables.   
   \begin{figure}
 \includegraphics[width=7.0cm]{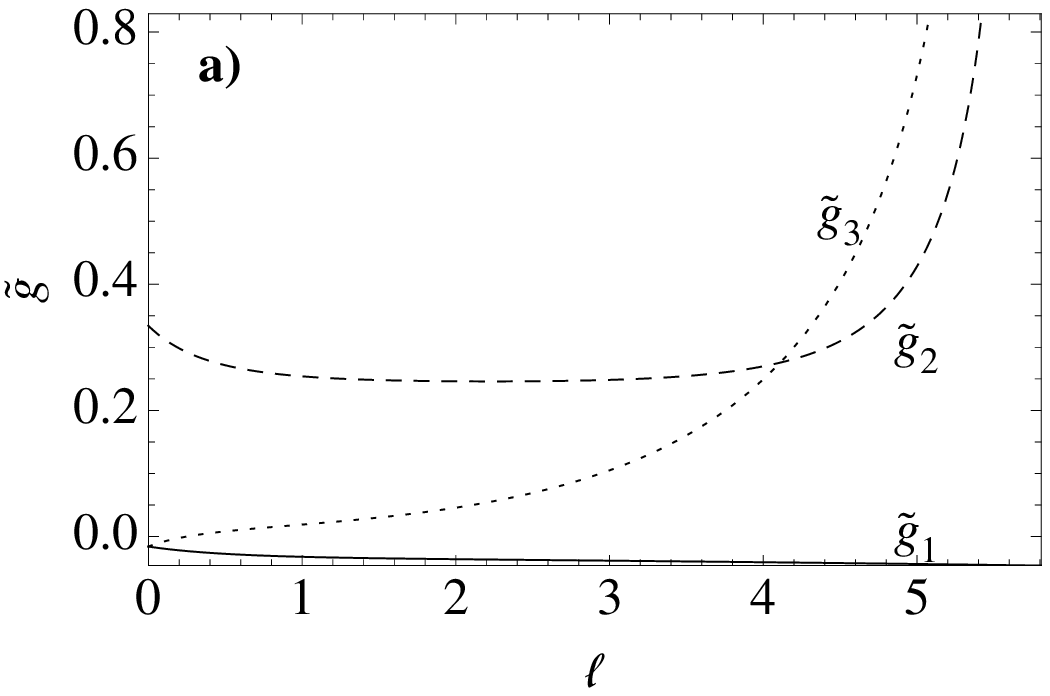} \includegraphics[width=7.0cm]{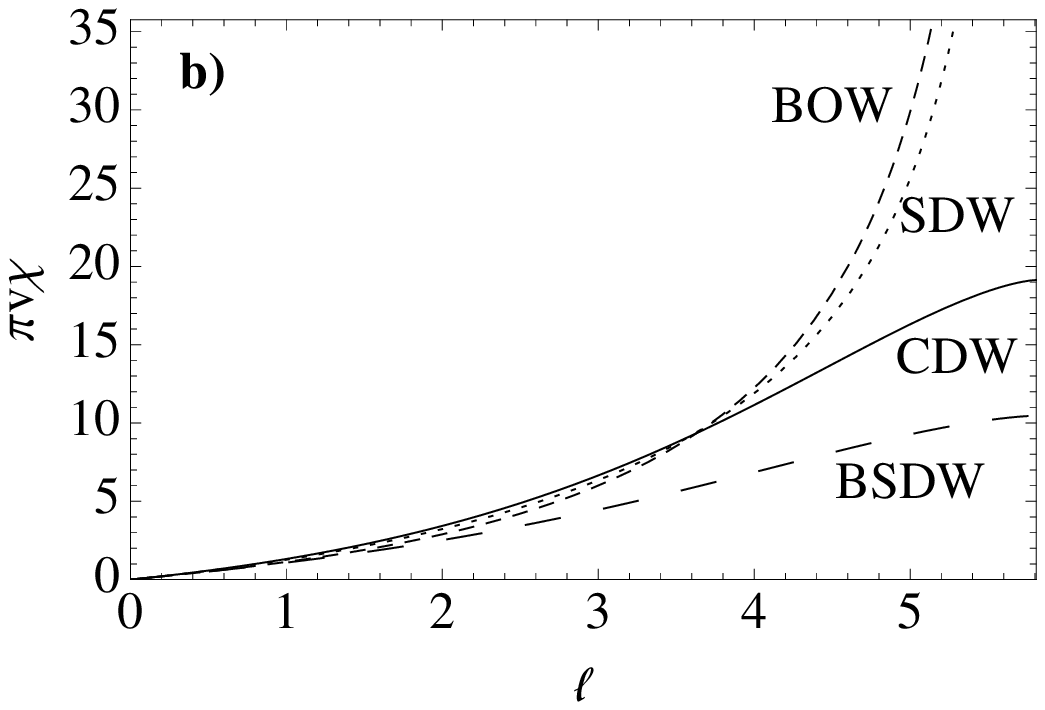}
 \caption{ a) Flow of the $\tilde{g}_{1,2,3}$ couplings at C  (1, 0.55) in the phase diagram of Fig.~\ref{Dphases}. b) The susceptibilities {\it vs} $\ell$;   \label{C}} 
 \end{figure}

\begin{figure}
 \includegraphics[width=7.0cm]{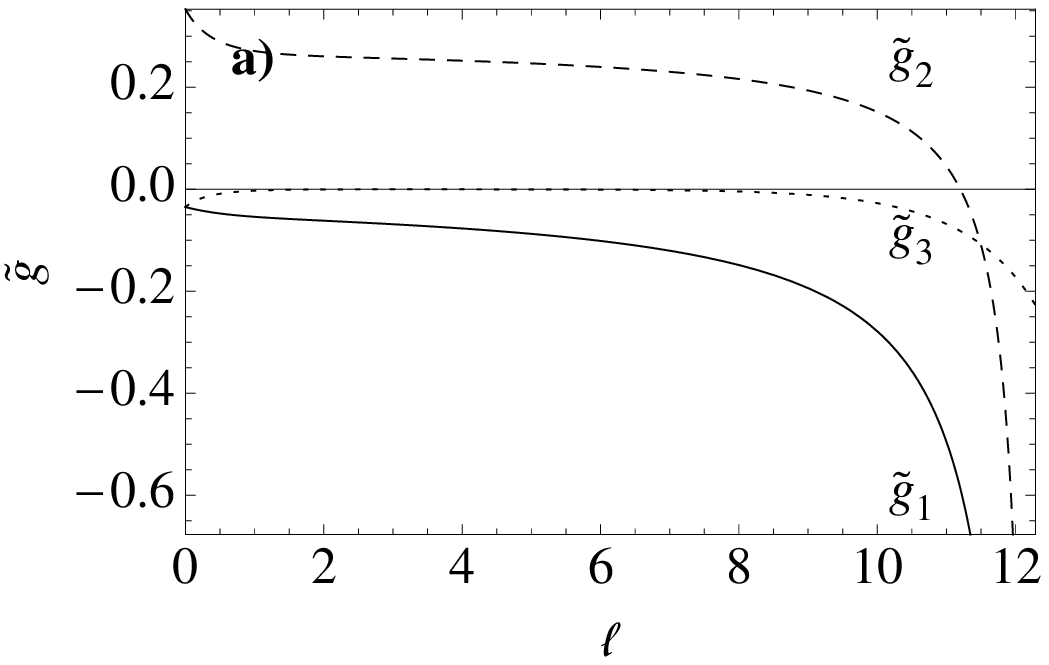} \includegraphics[width=7.0cm]{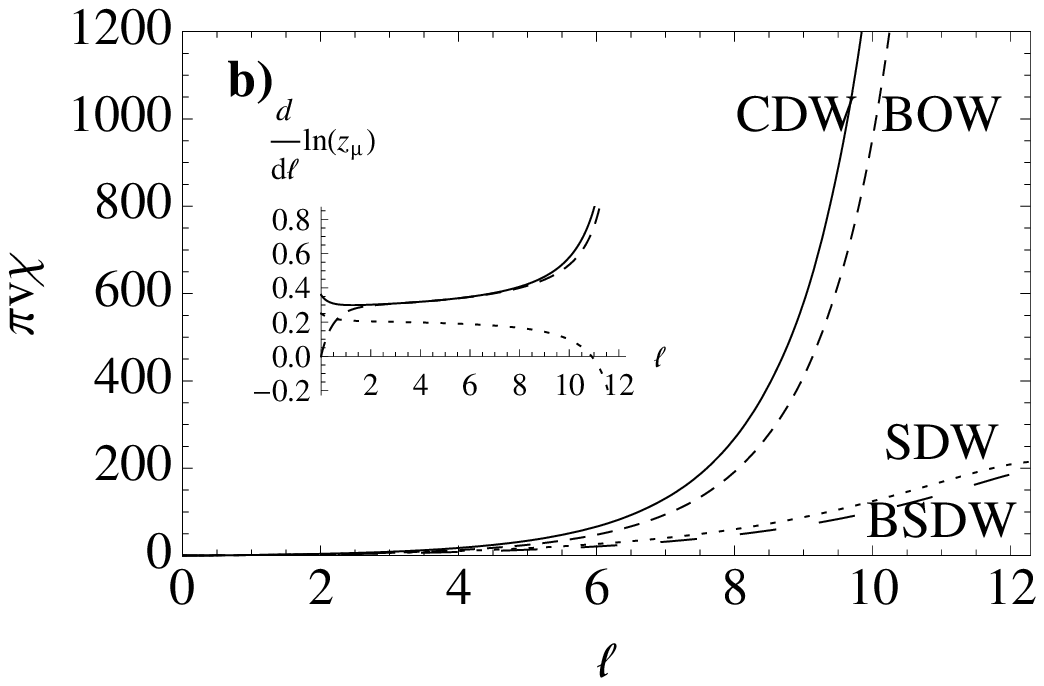}
 \caption{ a) Flow of the $\tilde{g}_{1,2,3}$ couplings at D  (1, 0.609) in the phase diagram of Fig.~\ref{Dphases}. b) The susceptibilities {\it vs} $\ell$; inset: the flow of the pair vertices $d_\ell \ln z_\mu$  for $\mu=$ CDW, BOW,  and SDW \label{D}} 
 \end{figure}

The dominance of the BOW phase becomes more pronounced as one moves up across  the line $U=2V$   (point C of Fig.~\ref{Dphases}-b). In this region,  the initial local couplings $\tilde{g}_1$ and $\tilde{g}_3$  are negative, but  the latter interaction is still pushed to strong repulsive sector by non-local couplings (Fig.~\ref{C}-a).  The BOW susceptibility  then develops the strongest singularity with the largest amplitude  (Fig.~\ref{C}-b). These features of the flow and the predominance of BOW order keep  on up the BOW-CDW boundary passing just below point D in Fig.~\ref{Dphases}-b.   At that point, strong attractive coupling in $\tilde{g}_1$ and $\tilde{g}_2$ is occurring while $\tilde{g}_3$ remains small and attractive (Fig.~\ref{D}-a), implying  the formation of a gap in the spin sector instead of the charge. In these conditions, we have $\tilde{g}_{\rm CDW}$ $>$  $\tilde{g}_{\rm BOW}$, which marks the onset of a dominant CDW phase.  The BOW order is subdominant and SDW correlations are non longer singular and are strongly reduced by the presence of a spin gap. 
 It is worth noting that   the emergence of a spin gap  regime  on the BOW-CDW frontier    is compatible with the results of quantum Monte Carlo simulations\cite{Sandvik04}, which find the onset of a Luther-Emery liquid with a spin gap at the boundary. 
The same analysis carried out as a function of $U$   allows for  the delimitation of  a small but finite fan-shape region of the weak coupling phase diagram of Fig.~\ref{Dphases} where the BOW order intervenes as the ground state  around the $U=2V$ line.  This well known result of numerical calculations \cite{Nakamura99,Nakamura00,Sandvik04} and functional RG\cite{Tam06} contrasts with the direct SDW to CDW transition predicted for the 1D-EG model\cite{Emery79}.
   \begin{figure}
 \includegraphics[width=7.5cm]{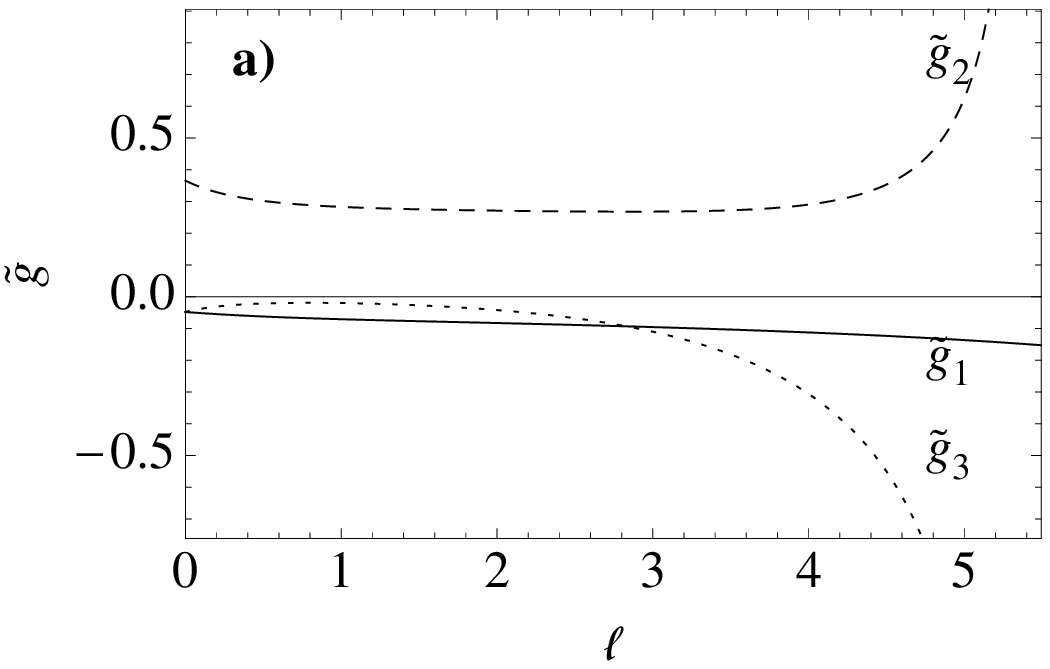}\par\hskip 0.3 truecm \includegraphics[width=7.0cm]{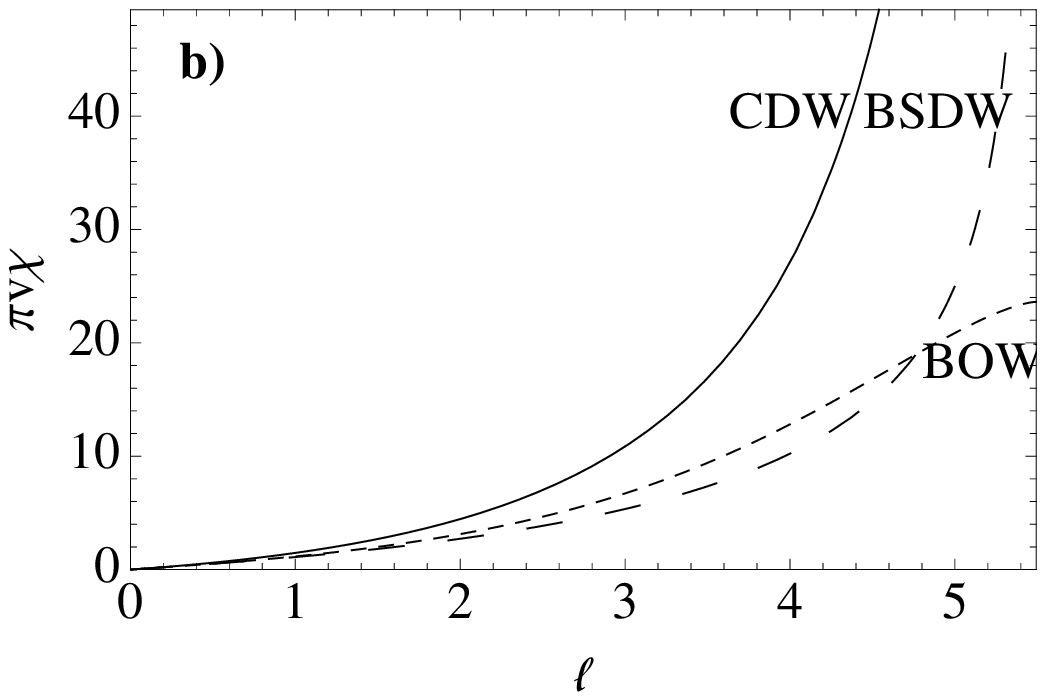}
 \caption{ a) Flow of the $\tilde{g}_{1,2,3}$ couplings at E  (1, 0.65) in the phase diagram of Fig.~\ref{Dphases}. b) The susceptibilities {\it vs} $\ell$. \label{E}} 
 \end{figure}

We proceed on  the analysis of the repulsive $U$ region by    looking at the point E, that is above the intermediate BOW  region. In this domain,   $\tilde{g}_2$ and $\tilde{g}_3$   scale to strong repulsive and attractive couplings, respectively, while $\tilde{g}_1$ is non universal and weakly attractive (Fig.~\ref{E}-a), contrary to what is found for the electron gas model\cite{Kimura75,Emery79,Solyom79}. The CDW  singularity is  stronger and    accompanied by a weaker singularity in the BSDW response (Fig.~\ref{E}-b). The BSDW replaces BOW as the subdominant phase over a finite  domain of the  phase diagram at $V>0$ (Fig.~\ref{Dphases}-a).

   \begin{figure}
 \includegraphics[width=7.0cm]{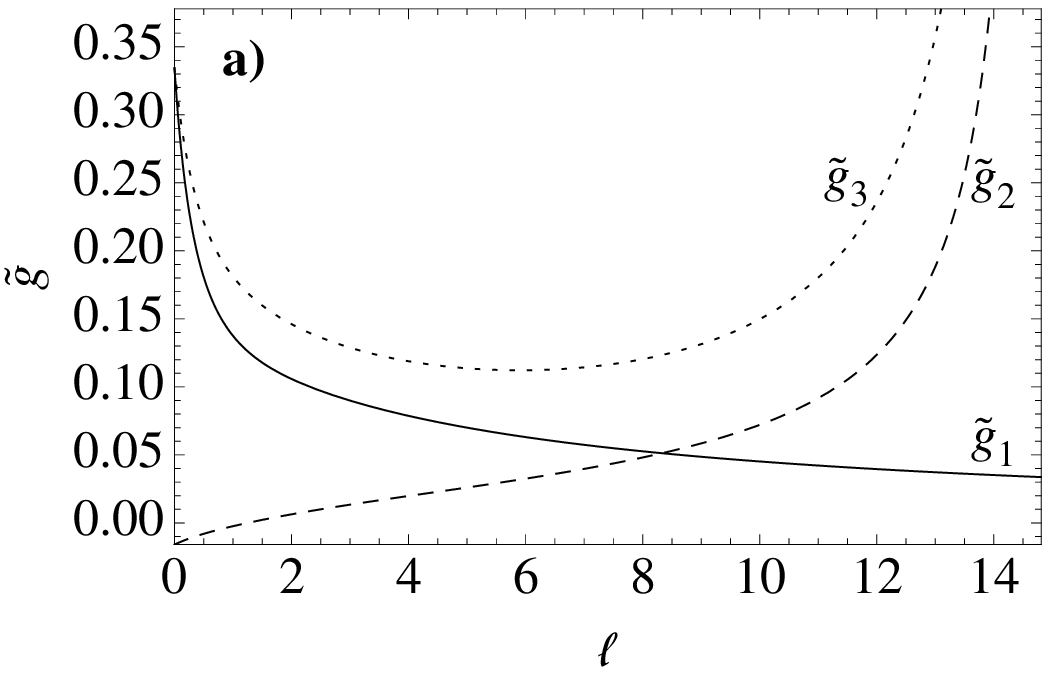}  \includegraphics[width=7.0cm]{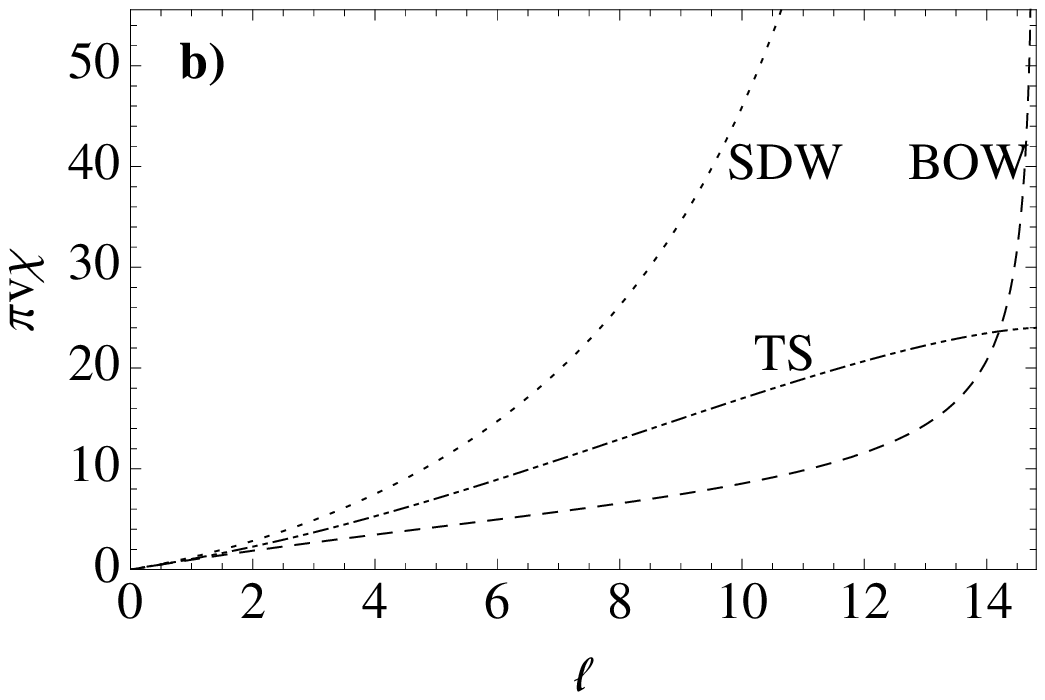}
 \caption{ a) Flow of the $\tilde{g}_{1,2,3}$ couplings at F (1, -0.55) in the phase diagram of Fig.~\ref{Dphases}. b) The susceptibilities {\it vs} $\ell$.   \label{F}} 
 \end{figure}
 We turn to  the   point F located in the $V<0$ region below, but near the $U=-2V$ line, where qualitative changes with respect to the continuum limit are also found. In the framework of the 1D-EG model, the region below the $U=-2V$ line  is characterized by the conditions $\tilde{g}_1 >0$  and $ \tilde{g}_1-2\tilde{g}_2 >\tilde{g}_3$, respectively  for gapless  spin and  charge degrees of freedom with dominant TS and subdominant SS phases\cite{Emery79,Solyom79}. In the presence of non local couplings, however, while $\tilde{g}_1$ is marginally irrelevant, both $\tilde{g}_2$ and $\tilde{g}_3$ scale  to strong repulsive coupling signaling that the charge degrees of freedom are still gapped (Fig.~\ref{F}-a). Therefore the SDW phase remains dominant   contrary to the 1D-EG prediction of a gapless TS phase \cite{Emery79,Solyom79,Kimura75} (Fig.~\ref{F}-b); the SDW incursion below the $U=-2V$ line expands in size  as $U$ increases  as shown in Fig.~\ref{Dphases}-a. It is worth mentioning that the resulting inward bending of the TS-SDW boundary line which becomes more pronounced with increasing $U$ is consistent with the numerical results of Nakamura \cite{Nakamura00}.

\begin{figure}
 \includegraphics[width=7.0cm]{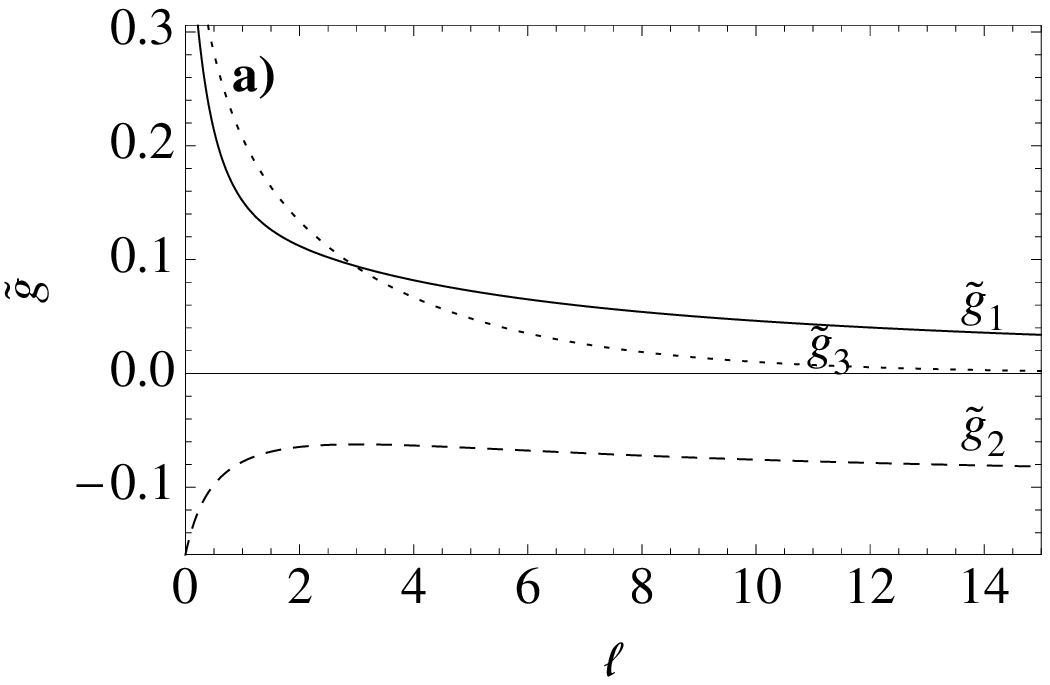}  \includegraphics[width=7.0cm]{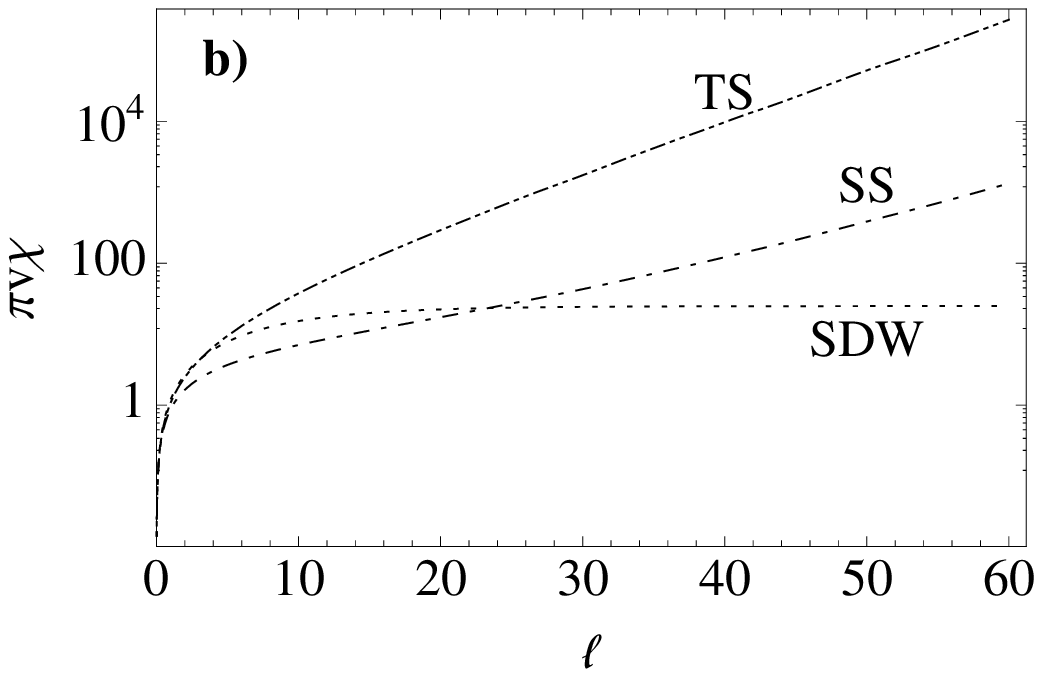}
 \caption{ a) Flow of the $\tilde{g}_{1,2,3}$ couplings at G (1, -1) in the phase diagram of Fig.~\ref{Dphases}. b) The susceptibilities {\it vs} $\ell$.  \label{G}} 
 \end{figure}

  Finally, as one moves sufficiently downward along the $V$ axis, one reaches a region where  $\tilde{g}_1$ and $\tilde{g}_3$ behave the way marginally irrelevant variables do (Fig~\ref{G}-a), as shown for instance at the point G of   the phase diagram of Fig.~\ref{Dphases}-a. One then essentially recovers the behavior of the 1D-EG model with a dominant (subdominant) power law singularity $\chi_{\rm TS(SS)} \propto \exp(\gamma_{\rm TS(SS)}\ell)$  for  TS (SS) response at large $\ell$  (Fig~\ref{G}-b) with $\gamma_{\rm TS} \gtrsim \gamma_{\rm SS}>0$.
\subsection{Attractive U } 
\begin{figure}
 \includegraphics[width=7.0cm]{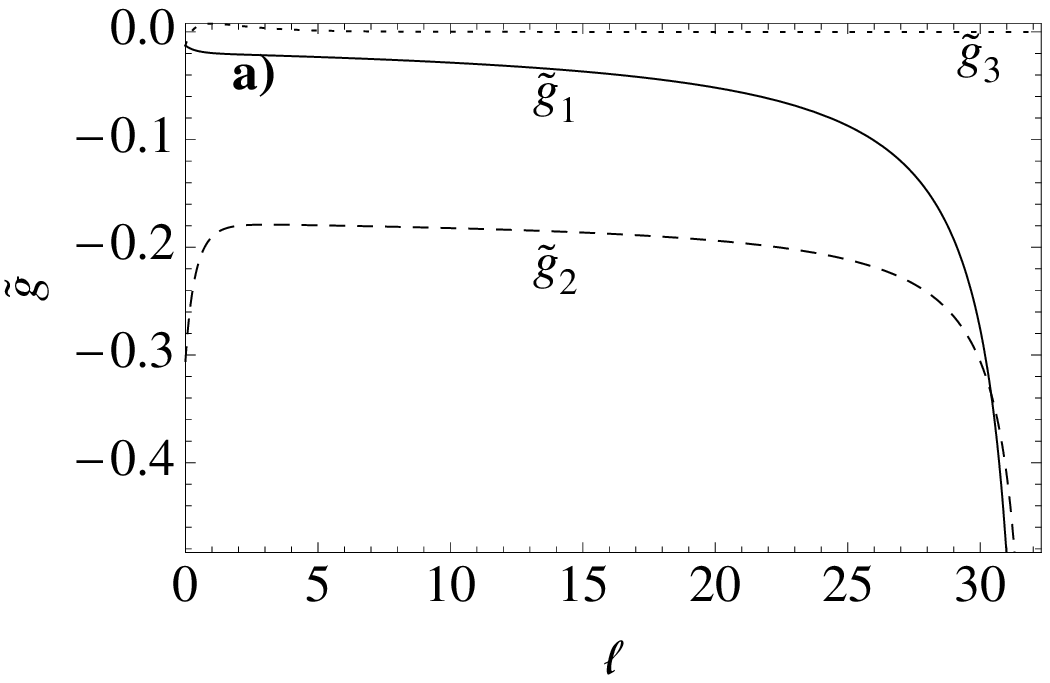}  \includegraphics[width=7.0cm]{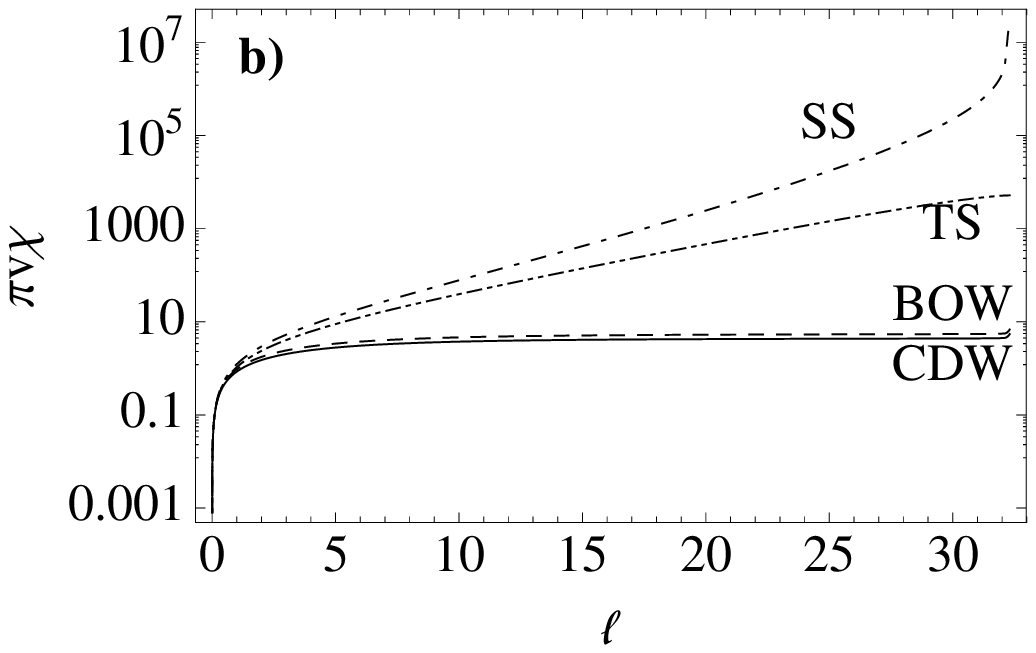}
 \caption{ a) Flow of the $\tilde{g}_{1,2,3}$ couplings at H (-1, -0.46) in the phase diagram of Fig.~\ref{Dphases}. b) The susceptibilities (logarithmic scale) {\it vs} $\ell$ . \label{H} } 
 \end{figure}
We now consider the region of negative $U$    near the $U= 2V$ line. In this region, we encounter an alteration of the  1D-EG phase diagram boundary that is similar to the  one discussed in the last paragraph at $U=- 2V>0$. At H in Fig.~\ref{Dphases}-b, a portion  of the phase diagram  with dominant (subdominant) TS (SS) gapless  phase is lost, this time to the benefit of a    SS phase with  a spin gap. Strong attractive coupling in the spin sector is induced by non local couplings that push downward the renormalization of $\tilde{g}_1$ (Fig.~\ref{H}-a). As for  Umklapp scattering,  it stays  weakly attractive indicating that the charge sector is gapless. The  SS-TS boundary is then distorted inward compared to the straight line  1D-EG prediction, which is in fair agreement with the results of exact diagonalisation by Nakamura.\cite{Nakamura00} The SS  phase expands from the bent boundary with the TS phase up to the $V=0$ symmetry  line for the transition to CDW (Fig.~\ref{Dphases}-a). We    exemplify the SS region by the point H of the phase diagram (Fig.~\ref{Dphases}-a), where the $\tilde{g}_1$ and $\tilde{g}_2$ scale   to strong attractive  coupling for the formation of a spin gap at $ \ell_\sigma$ (Fig.~\ref{H}-a). The SS response is the only singular response of the system and the whole region has   no subdominant phase (Fig.~\ref{H}-b).    
\begin{figure}
 \includegraphics[width=7.0cm]{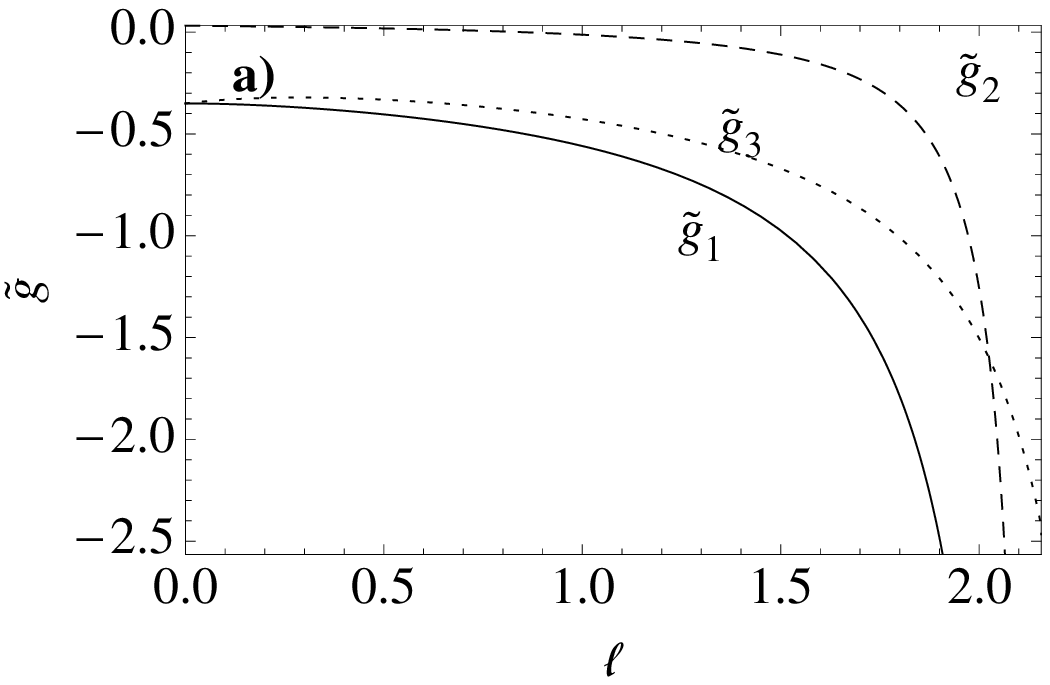}  \includegraphics[width=7.0cm]{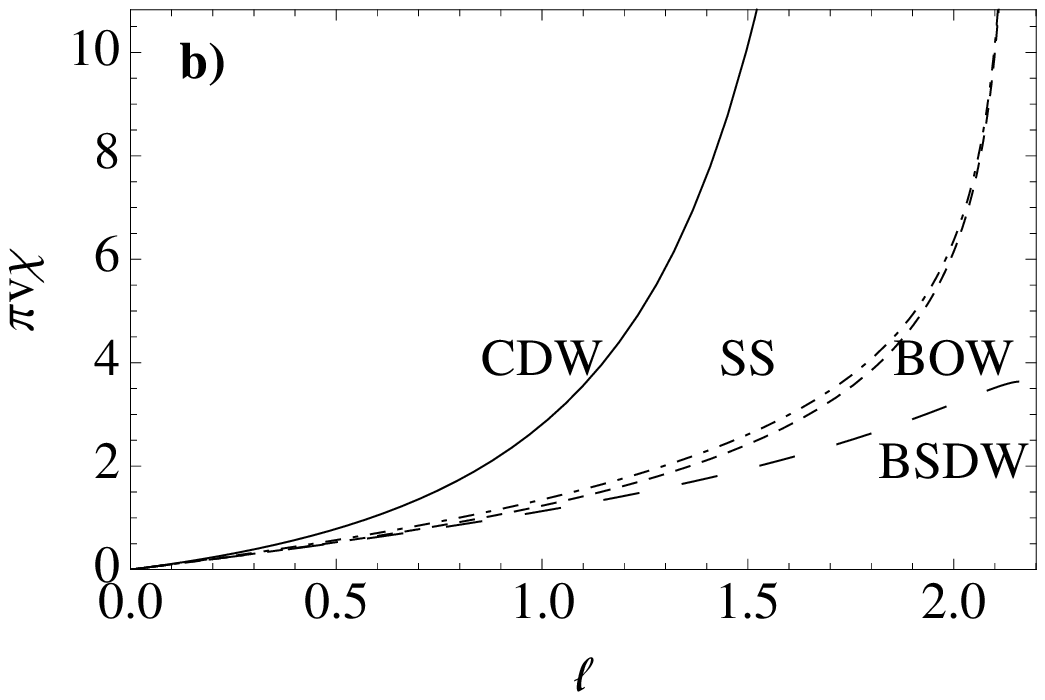}
 \caption{ a) Flow of the $\tilde{g}_{1,2,3}$ couplings at I (-1, 0.6) in the phase diagram of Fig.~\ref{Dphases}. b) The susceptibilities {\it vs} $\ell$.  \label{I} } 
 \end{figure}

We end the tour of the phase diagram with the second quadrant   above the $V=0$ SS-CDW frontier at the point I. There, the rapid flow to strong attractive coupling for  $\tilde{g}_1$ marks the onset  of a spin gap  at relatively small $\ell_\sigma$ (Fig.~\ref{I}-a). The strong attraction for $\tilde{g}_1$  prevails over the Umklapp term,  though also marginally relevant. The  singularity of the CDW response is thus by far  prevalent, being followed by a much weaker  BOW  susceptibility, whose subdominance is  less guaranteed  since  it occurs in the strong coupling domain where the perturbative RG becomes  less reliable.

 
 \section{Conclusion}
 In this work we have proposed a generalization of the  momentum shell  renormalization group transformation that is applicable to 1D lattice models of interacting electrons. The approach has been put to the test for the determination of the phase diagram of the extended Hubbard model in weak coupling. The  method discloses  the influence of a finite number of dangerous irrelevant couplings on the scaling of marginal interaction terms  of the  model.  Modification of scaling gives rise in some regions of the phase diagram to   unexpected phases from the standpoint of the  theory in the continuum limit.  Among the results  obtained,   let us mention the incursion of  BOW order in a finite portion of the repulsive $U\simeq 2V$ sector of the phase diagram, which agrees with previous  results of  numerical and functional RG methods.  The approach is also able  to capture  the deformation of  boundaries between Luttinger liquid and  gapped phases in the phase diagram of the model as  found previously 
 by exact diagonalisation. 
 
 These   findings   are encouraging for   applications  to other weak coupling   1D or quasi-1D interacting electron models in which  lattice details can  play an important role in the  properties of correlations at long distance. 
 
 \acknowledgments
 C. B thanks the National Science and Engineering Research Council  of Canada (NSERC), the R\'eseau Qu\'ebcois des Mat\'eriaux de Pointe (RQMP) and  the {\it Quantum materials} program of Canadian Institute of Advanced Research (CIFAR) for financial support.
  \bibliography{/Users/cbourbon/Dossiers/articles/Bibliographie/articlesII.bib}
\end{document}